\documentclass[sigconf]{acmart}
\usepackage{booktabs} 

\usepackage{seqsplit}

\usepackage{amsmath,epsfig}
\usepackage{graphicx}
\usepackage{caption}
\usepackage[font=small]{caption}
\usepackage{subfig}
\usepackage{wrapfig}
\usepackage{lipsum}
\usepackage{algorithm}
\usepackage{algpseudocode}
\usepackage{amssymb}
\usepackage{pifont}

\algtext*{EndWhile}
\algtext*{EndIf}
\algtext*{EndFor}

\usepackage{setspace}
\usepackage{color,soul} 

\usepackage{soul}
\usepackage{color}

\renewcommand\footnotetextcopyrightpermission[1]{} 
\begin{document}
\title{Using Multi-Core HW/SW Co-design Architecture for Accelerating K-means Clustering Algorithm}

\author{Hadi Mardani Kamali}
\affiliation{%
  \institution{Department of Electrical and Computer Engineering}
  \city{George Mason University}
}
\email{hmardani@gmu.edu}

\renewcommand{\shorttitle}{}
\renewcommand{\shortauthors}{}

\begin{abstract}
The capability of classifying and clustering a desired set of data is an essential part of building knowledge from data. However, as the size and dimensionality of input data increases, the run-time for such clustering algorithms is expected to grow superlinearly, making it a big challenge when dealing with BigData.

K-mean clustering is an essential tool for many big data applications including data mining, predictive analysis, forecasting studies, and machine learning. However, due to large size (volume) of Big-Data, and large dimensionality of its data points, even the application of a simple k-mean clustering may become extremely time and resource demanding. Specially when it is necessary to have a fast and modular dataset analysis flow. 

In this paper, we demonstrate that using a two-level filtering algorithm based on binary kd-tree structure is able to decrease the time of convergence in K-means algorithm for large datasets. The two-level filtering algorithm based on binary kd-tree structure evolves the SW to naturally divide the classification into smaller data sets, based on the number of available cores and size of logic available in a target FPGA. The empirical result on this two-level structure over multi-core FPGA-based architecture provides 330$\times$ speed-up compared to a conventional software-only solution. 
\end{abstract}

%
%

\copyrightyear{2018} 
\acmYear{2018} 
\setcopyright{acmcopyright}
\acmConference[GLSVLSI '18]{2018 Great Lakes Symposium on VLSI}{May 23--25, 2018}{Chicago, IL, USA}
\acmBooktitle{GLSVLSI '18: 2018 Great Lakes Symposium on VLSI, May 23--25, 2018, Chicago, IL, USA. }
\acmPrice{15.00}
\acmDOI{10.1145/3194554.3194648}
\acmISBN{978-1-4503-5724-1/18/05}


\maketitle

\vspace{-6pt}
\section{Introduction}

Cloud computing when combined with the Internet of Things has enabled may new services such as Sensing As a service (SENaaS), Sensor Data as a Service (SDaaS) and Sensor Trigger as a Service (STaaS). The applications that use such services may be of predictive, data mining, machine learning, or forecasting nature \cite{Jain2011, Theiler1997, Estlick2001, Roshanisefat2018srclock, kamali2018lutlock}, many of which are in need of categorizing and clustering of very large and high-dimensional input data-sets as a part of their larger computational flow. The ability to classify and cluster a desired set of data to see trends, similarities, correlations, and trajectories is an essential part of building knowledge from data. However, as the size and dimensionality of input data increases, the run-time for such clustering algorithms is expected to grow superlinearly, making it a big challenge when dealing with BigData \cite{Sayadi2018cf, Sayadi2017iccd}.

The goal of clustering is to classify the data according to a specific metric such that objects within a cluster/group, in terms of having a feature, fitting a description, or displaying a characteristic are similar, while they are different from the members that are located in other groups. 

There are different categories for clustering. A clustering algorithm may be supervised (hierarchical) or unsupervised (un-nested). It may be exclusive or fuzzy, and could be complete or partial. Depending on the type of clustering algorithms used, the resulting clusters may be well separated, prototype-based (centroid-based), graph-based, or density-based \cite{Tan2013}.

K-means is one of the simplest and yet most used \cite{Tan2013, Wu2012} centroid-based unsupervised clustering algorithms. Although categorized as simple and low complexity classification function, its applicability to large data-sets (Big-Data in general) depends on the scalability of its software (SW) implementation with respect to the available hardware (HW). One of the most promising HW platforms that is leveraged for achieving considerable speedup in big data applications are FPGAs. Recent FPGAs are equipped with hundreds of thousands of fine-grained logics and coarse-grained communications, which provide huge parallelism with negligible communication cost. FPGA solutions enable higher parallelism than clusters of CPUs or GPUs at a much lower cost, but with greater mapping overhead. However if the size of data is large, such that the mapping-time overhead is small or negligible compare to the run-time of the targeted application (such as bigData clustering), using FPGA-based solutions are preferred.

When it comes to comparison with ASIC accelerators, the FPGA solutions, in terms of power and performance, are not as efficient. However, they could be re-purposed from application to application, where as ASIC accelerator maintains a fixed behaviour. Hence, FPGAs are a better solution for general purpose computing environment, such as cloud data-centers, where the applications are dynamic and priori-unknown \cite{Sayadi2018aspdac, Sayadi2018dac}. In such cases, in which dynamic process is more preferred, the high cost of custom ASIC accelerators are not well justified, and the re-configurability and adaptability of FPGA-accelerated solutions are greatly desired. 

In order to provide improvement for the usability of FPGA solutions in dealing with semi-parallel applications, the FPGAs are equipped with mid to high-performance multi core processors (e.g. ARM Cortex A9, A12, A15). The existence of multiple mid to high-performance cores on the same die as FPGA improves the efficiency of HW/SW co-design \cite{kamali2018ducnoc, kamali2016adapnoc, kamali2018swift, zynq7000} and provide greater flexibility is using FPGAs in data centers as re-configurable, yet powerful hardware accelerators \cite{Freund2016, Wilson2014}. 

The execution time of k-clustering algorithms could be improved by means of both SW and HW. In fact, using software-based techniques and methods provides performance improvement in case of k-means algorithm. For example, on the SW side, one could use (1) binary kd-tree structure for dividing search space members into "boxes" \cite{Kanungo2002}, and (2) triangle inequality for avoiding redundant distant calculation \cite{Elkan2003}. On the other hand, using hardware-based architectures, like FPGA-based implementation, accelerate the algorithm considerably. on the HW side, more capable or additional computing resources reduce the computational time. For example, by directly mapping a k-means clustering algorithm to a capable FPGA, a considerable reduction in execution time, compared to a sole SW-based solution, is expected. This speed-up is the result of throwing additional hardware to speed up the parallel kd-clustering algorithm. However, such direct and non-optimized mapping of software intended for CPUs to FPGAs does not result in best utilizing all FPGA resources. Hence, to maximize the FPGA utilization, and to speed up non-parallel portions of the code, a more precise SW/HW co-design is required \cite{Sayadi2017igsc, Roshanisefat2018bench}.

In this paper, we demonstrate that using a HW/SW co-design architecture as well as a software-based technique, i.e. kd-tree clustering algorithm, considerably reduces the execution time of k-means algorithm. For this purpose a mapping and an aggregation function have been implemented on top of kd-tree clustering algorithm. This approach allows us to divide the work across the hardware, i.e. logic and multiple cores in an FPGA, to gain the maximum achievable speedup by utilizing all available resources. Additionally, we demonstrate that having a custom high-performance DMA for transmitting data between host and FPGA via PCI Express (PCIe) interface, significantly reduces the execution time overhead related to data transmission time, and provides better speedup in comparison with a conventional software based solutions. 

The rest of the paper is organized as follows. The k-means theory and algorithm are described in Section 2. Section 3 briefly illustrates structure of binary kd-tree for filtering algorithm. The architecture of HW/SW co-design architecture is elaborated in Section 4. Experimental results are shown in Section 5. Section 6 covers the related work. And finally, Section 7 concludes the paper.

\section{K-Means Clustering Algorithm}

K-means is one of the simplest partitioning algorithms with fast execution time, which is popular for unsupervised centroid-based clustering. As it names implies, k-means divides input datasets to "k" groups, called clusters, where all members in a cluster are similar in some metrics, and they are dissimilar to members of other clusters. Additionally, k-means is a centroid-based algorithm, where each cluster has a prototype which is indicator of the cluster. Each data point will be classified into a cluster whose centroid is the closest. Three conventional distance metrics have been used for k-means clustering to calculate the distance between each point and centroids: \emph{Manhattan}, \emph{Max}, and \emph{Euclidean} \cite{Estlick2001}. For instance, if we suppose that each data point is a vector $\overrightarrow{dp} = (p_1, p_2, ..., p_m)$, the Euclidean distance can be defined as follow:
\begin{equation}
\vspace{-3pt}
EuclidDist(\overrightarrow{dp},\overrightarrow{cent}) = \big(\displaystyle\sum_{i=1}^{m} (dp_i - cent_i)^2\big)^{\frac{1}{2}}
\vspace{-1pt}
\end{equation}

The k-means algorithm first initiates $k$ centroids. Then it enters an iterative process where each iteration consists of two steps: (1) \emph{Assignment Step}: where each point will be assigned to a cluster whose centroid is the closest. (2) \emph{Update Step}: where a new centroid is found by re-calculating the mean of new assigned points to each cluster. When the centroids stop changing, the clustering of datasets to $k$ clusters is successfully accomplished.

\vspace{-1pt}
\section{Binary kd-Tree for Filtering Search Space}

The filtering algorithm is developed based on a binary kd-tree, which reduces the required time for search queries \cite{Kanungo2002}. In this algorithm, all data points recursively will be divided into some axis-aligned bounding boxes. Also, hierarchical divisions are accomplished based on \emph{axis-aligned bounding boxes}. This recursive process generates a multi-dimensional binary search tree, whose root is a bounding box of all data points, and each level of the tree consists of two meaningful subsets of data points, and consequently each leaf represents at most one data point. Each node stores some essential information, such as the corresponding bounding box members (\emph{cell}), the number of data points in the box (\emph{count}), the weighted centroid (\emph{wgtCent}) which represent the sum of all data points in a box, and candidates for centroid (\emph{Z}).

\begin{algorithm}
\caption{Filtering algorithm by using binary kd-Tree \cite{Kanungo2002}}\label{Filter}
\begin{algorithmic}[1]
\small
\Function{Filter}{kdNode~$u$, CandidateSet $Z$}
\State $C\gets u.cell$;
\If {$u$~is~a~leaf}
    \State $z^*\gets the~closest~point~in~Z~to~u.point$;
    \State $z^*.WgtCent\gets z^*.WgtCent~+~u.point$;
    \State $z^*.count\gets z^*.count~+~1$;
\Else
    \State $z^*\gets the~closest~point~in~Z~to~C's~midpoint$;
    \ForAll{$z \in Z \setminus \{z^*\}$}
         \If {$z.isFather(z^*,C)$}
              \State $Z\gets Z \setminus \{z^*\}$;
         \EndIf
    \EndFor	  
    \If{$| Z | == 1$} 
        \State $z^*.WgtCent\gets z^*.WgtCent~+~u.wgtCent$;
        \State $z^*.count\gets z^*.count~+~u.count$;
    \Else
	  \State Filter($u.left$, $Z$);
        \State Filter($u.right$, $Z$);
    \EndIf
\EndIf
\EndFunction
\end{algorithmic}
\end{algorithm}

Alg. \ref{Filter} depicts the kd-tree filtering algorithm. In each node, \emph{Z} determines a subset of candidate centroids. If we suppose that we have $k$ clusters, the candidate centroids for the root node are all $k$ centroids, and the candidates for each internal node are a subset of $k$ clusters. Additionally, the candidates of each node are the nearest neighbors for some points. For each node, the distance between $z^* \in Z$ and the midpoint of \emph{cell} is calculated and compared with $z \in Z\setminus z^*$. Then, the sub-tree at the larger distance is pruned.

\vspace{-1pt}
\section{HW/SW co-design Two-level K-Means clustering}

\begin{figure}[t]
\centering
\includegraphics[width = 220pt]{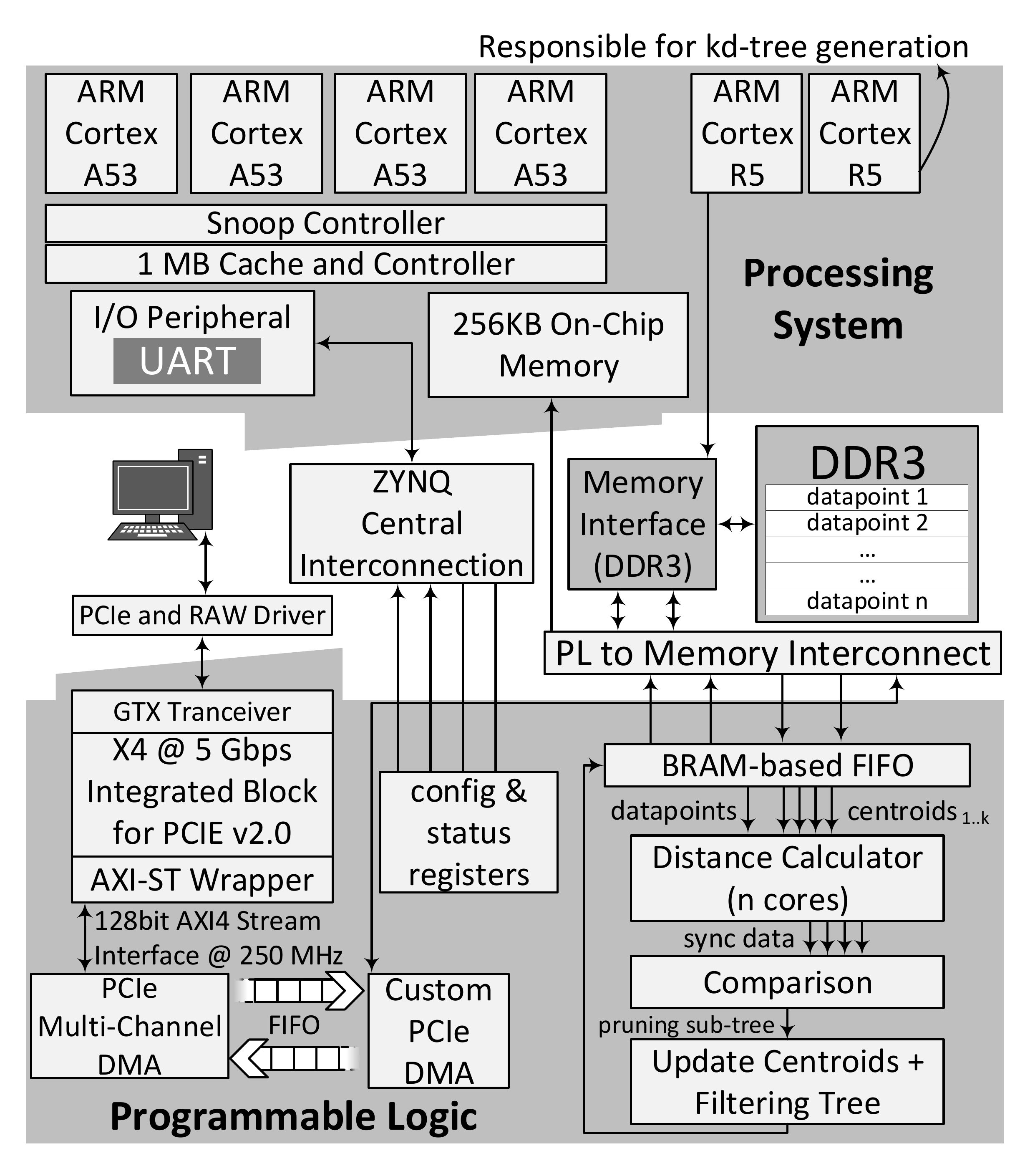}
\vspace{-10pt}
\caption{Overall MUCH-SWIFT System-Level Architecture.}
\label{TopArch}

\end{figure}

Similar to \cite{Estlick2001, Gokhale2003, Choi2014, Abdelrahman2016, Canilho2016}, we demonstrate that MUCH-SWIFT \cite{kamali2018swift}, as a HW/SW co-design architecture, accelerates k-means algorithm by using a system-level architecture, which consists of multiple processors and a single FPGA. A ZCU102 evaluation board has been used in this architecture, which is equipped with a Zynq-7000 Ultrascale+ SoC which is applicable for multi-core architectures. This architecture consists of two major sub-modules: (1) Processing System (PS) which consists of a Quad Cortex-A53 processor and a dual Cortex-R5 co-processor, (2) Programmable Logic (PL) which is responsible for implementing the parallel arithmetic cores that are required for Manhattan distance calculations, comparators, and updater. Fig. \ref{TopArch} illustrates the overall architecture of MUCH-SWIFT. As illustrated, it is implemented based on ZYNQ Ultrascale+ architecture, which has four Cortex-A53 up to 1.5 GHz, two Cortex-R5 up to 600 MHz, 1 GB DDR3 off-chip memory, and ZU9EG FPGA chip with around 600K logic cells. To achieve the highest speedup All processors are employed in this design; Each Cortex-A53 core is responsible for evaluating and analyzing one quarter of data points independently. Additionally, in order to reduces the search time, a binary kd-tree structure to filter (prune) some nodes and their children has been employed, which their candidates are not the nearest centroid. Then, in order to maximize the utilization of all four Cortex-A53 cores, N parallel clusters (N being the number of available cores, which is 4 in the experimental results section) has been built by dividing the original data-set into N smaller data-sets at the top of the kd-tree. After clustering the sub-data-sets, they are merged together and the filtering algorithm is invoked on top of the merged clusters. Using this two-layer clustering approach increases convergence, which decrease the required iterations for clustering. Furthermore, MUCH-SWIFT is able to process large datasets as well as large data size by using a DMA-based PCIe interface and DDR3 memory in ZCU102 without any significant throughput degradation. One of Cortex-R5 is responsible for handling the custom DMA between PCIe and DDR3 memory, and another Cortex-R5 must generate initial states of each quarter of data points as well as initial values of centroids. Also, controlling the update procedure after pruning in kd-tree structures, and update stage for centroids are accomplished by the second Cortex-R5.

\vspace{-6pt}
\subsection{Parallelism in kd-tree Traversal}

In order to maximize the parallelism in MUCH-SWIFT architecture, each Cortex-A53 core is made responsible for a quarter of data points. In fact, according to the size of data points, they should be divided into 4 independent groups, and each group is considered as a separate dataset. So, there are four independent kd-tree structures for all quarters, and filtering algorithm can be accomplished on each structure independently. Therefore, all sub-modules for k-means clustering algorithm, including the distance calculator, comparison, and updater are parallel and dedicated for each group. 

The big challenge in this architecture is the combining of the results of four divided sub-datasets. In order to accomplish k-clustering by means of this technique, i.e. dividing into four groups of data, it seems that it is necessary to implement four k $\frac{k}{4}$-clustering algorithms separately, and then gather four $\frac{k}{4}$ centroids as well as their corresponding clusters to provide k clusters. But, since dividing the dataset into four sub-datasets changes the calculated centroids, the obtained results in this scenario are not equivalent with a conventional k-clustering, and consequently the results are invalid. Therefore, a two-layer clustering mechanism has been implemented in order to perform it accurately. In the first level of k-means clustering, the data points will be divided into four independent sub-datasets, but, k clusters will be calculated for each sub-group, and after completing k-means clustering for each sub-group, all 4k centroids and their clusters ($4k$ clusters) will be gathered. So, it is necessary to combine a cluster in each sub-group with three clusters in other sub-groups with the nearest centroids. After merging four sub-datasets, the centroids and cluster members must be updated. When using this process, the second level of k-clustering has initial values (i.e. centroids and their clusters) that are considerably close to the final result. In fact, the number of iterations for second level of k-clustering is very small. 

\begin{algorithm}
\caption{Two-level k-clustering Algorithm by Using 4 parallel Binary kd-Tree Structures}\label{ParallelClustering}
\begin{algorithmic}[1]
\small
\Function{ParallelClustering}{DataPoint\_Set~$DP$}
\For{$i=0$ to $3$}
    \State DataPoint\_Set$~QDP[i]\gets Quarter(DP,i)$;
    \State kdNode $~*kdu[i] \gets Gen\_KdTree(QDP[i])$;
    \State CandidateSet$~Z\_Update[i] \gets Lloyd\big[QDP[i]\big]$;
    \State CandidateSet$~Z\_Current[i] \gets Z\_Update[i]$;  
\EndFor
\For{$i=0$ to $3$} \Comment{parallel in PL}
    \State Filter($kdu[i]$, $Z\_Update[i]$);
    \While {$Z\_Update[i] \neq Z\_Current[i]$}
        \State $Z\_Current[i] \gets Z\_Update[i]$; 
        \State Filter($kdu[i]$, $Z\_Update[i]$);
    \EndWhile
\EndFor
\State kdNode$~ kdu\_top \gets Combine(kdu[0:3])$;
\While{$Z$ is updated}
    \State Filter($kdu\_top$, $Z$);
\EndWhile
\EndFunction
\end{algorithmic}
\end{algorithm}

Alg. \ref{ParallelClustering} illustrates the pseudo-code of MUCH-SWIFT method. During the initialization state, dataset is divided into four separate sub-datasets via Quarter function. then A kd-tree is generated for each sub-dataset, and the Lloyd function is employed for choosing initial centroids \cite{Lloyd1982, kamali2016aes, Sayadi2014dft}. The most important part of this algorithm is the parallelism in tree traversal (\emph{Line} 8-14), where each Cortex-A53 core is made responsible for transceiving data to/from PL in order to calculate and update its corresponding kd-tree characteristics (i.e. centroids and clusters) in parallel.

\vspace{-2pt}
\subsection{No Limit for Dataset Size via High Throughput DDR3 Memory}

DDR3 off-chip memory in ZYNQ Ultrascale+ has been employed to maximize the feasible size of data. ZYNQ Ultrascale+ provides an efficient and fast DDR3 memory, which is accessible from both PS and PL, illustrated in Fig. \ref{TopArch}. The capacity of this memory is 1 GB, and it has a 128 bits data-bus for read/write access. Also, as it can be seen in Fig. \ref{TopArch}), it is necessary to implement a BRAM-based bridge (\emph{BRAM-based FIFO}) between DDR3 and PL in order to transfer data from PL to DDR3 and vice versa. In order to minimize the required BRAM-based bridge size between DDR3 and PL, the data size for each level of tree traversal  has been evaluated separately. Similar to \cite{Winterstein2013}, hierarchical access provides this possibility to release and reuse the memory at each level (depth) before starting the next level (depth) of tree. In addition, all data is permanently kept in DDR3; hence, overwriting the data can be accomplished without any throughput degradation. As a result, the large size of DDR3 provides this possibility to maximize the dataset size as required. For instance, suppose that MUCH-SWIFT is configured to classify $N = 100000$ data into $K=1024$ clusters. In the worst case, the structure of kd-tree is like a \emph{degenerate} tree. In this case, we need $(N-1)\times(K)\times(log_2K) \simeq $~122 MB, which is much less than the DDR3 memory, i.e. 1GB.

\vspace{-2pt}
\section{Experimental Results}

In order to demonstrate the MUCH-SWIFT throughput, some test cases should be considered. A large Xilinx ZYNQ-based SoC architecture (ZCU102 evaluation board) has been targeted to evaluate this architecture. ZYNQ can facilitate software side development by using Xilinx SDK. Furthermore, Vivado 16.2 is used for synthesizing, implementing, and downloading the overall design on FPGA, which provides this possibility to implement a block diagram for all parts of the design, even software side. MUCH-SWIFT consists of four main sub-modules:

\begin{enumerate}

\item PS consists of a quad Cortex-A53 core and a dual Cortex-R5 core, which is responsible for controlling the transceiving data to/from each core (Cortex-A53) from/to PL in order to perform k-clustering computations. Also, one Cortex-R5 should handle custom DMA for transmitting data to DDR3 from PCIe interface, and other Cortex-R5 core controls the updating stage of the filtering algorithm.

\item All floating point arithmetic operations, i.e. Manhattan distance, compare, and update centroids have been accomplished in PL.

\item As illustrated in Fig. \ref{TopArch}, an UART interface has been engaged to determine the number of clusters as a configurable parameter. In fact, the number of clusters is used to determine the number of parallel modules in PL. For instance, if we set the number of clusters to $K = 5$, since there are four sub-datasets, and each sub-dataset should implement a ($K = 5$)-clustering, we will have 20($5\times4$) parallel modules, including Manhattan Distances, compares, and updates, to accomplish the computations. So, the number of clusters has been used as a configurable parameter for PL in order to generate the logic for parallel computation modules.

\item PCIe interface is employed for transmitting datasets from the host to PL. Note that all interconnections between top modules in MUCH-SWIFT architecture is implemented based-on AXI. a 128-bit AXI has been employed between PL and PS as well as between DDR3 and PS/PL in order to guarantee the required throughput. Also, a 64-bit AXI-based data-bus has been implemented to establish the custom DMA between PCIe and DDR3 efficient. PS is developed in C++ using Xilinx SDK, and PL is implemented in Verilog HDL using Xilinx Vivado. Also, all sub-modules in PL are implemented in AXI-based structure.

\end{enumerate}

\begin{figure}[t]%
    \centering
    \vspace{-7pt}
    \subfloat[]{{\includegraphics[width=240pt]{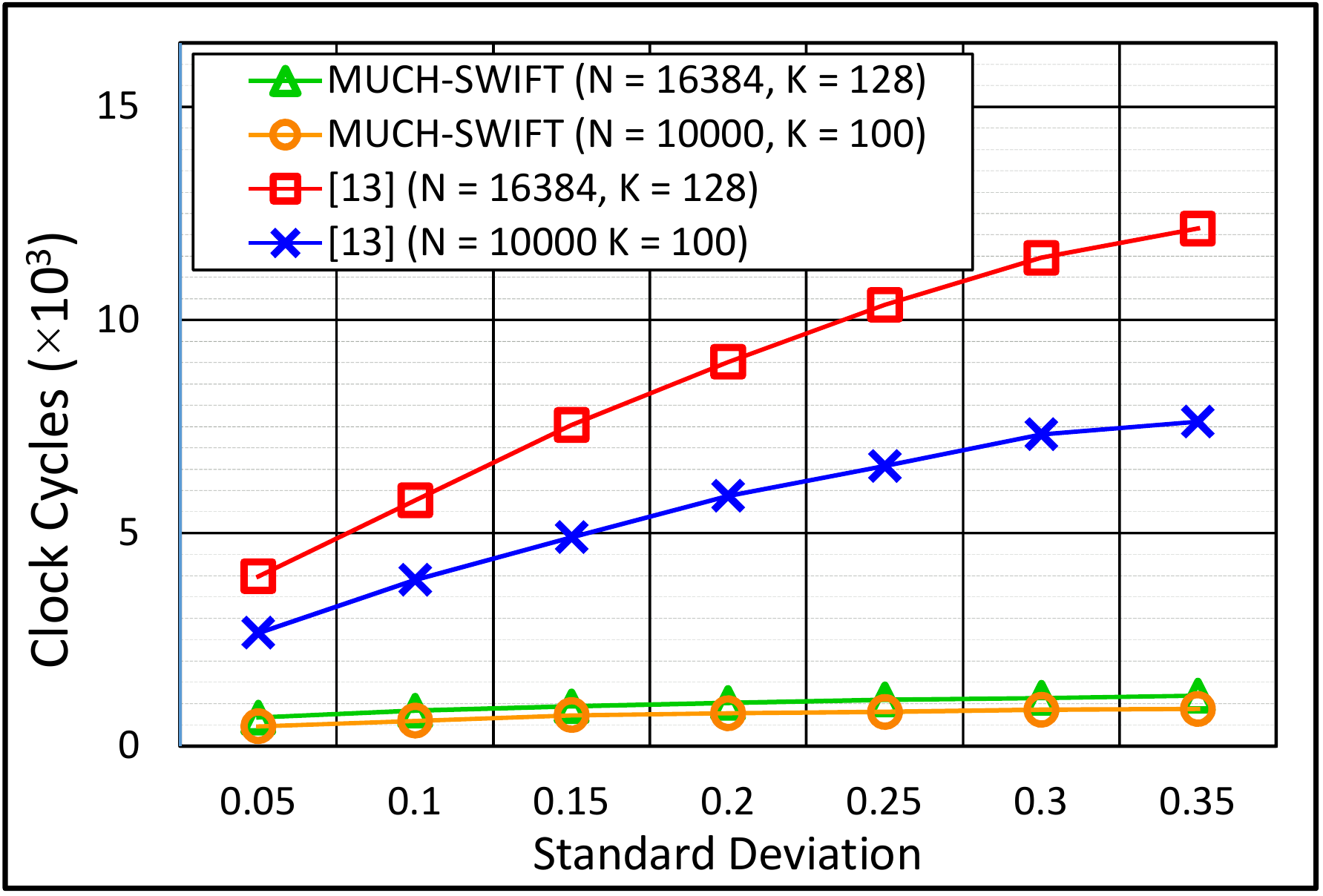} }} \\
    \subfloat[]{{\includegraphics[width=240pt]{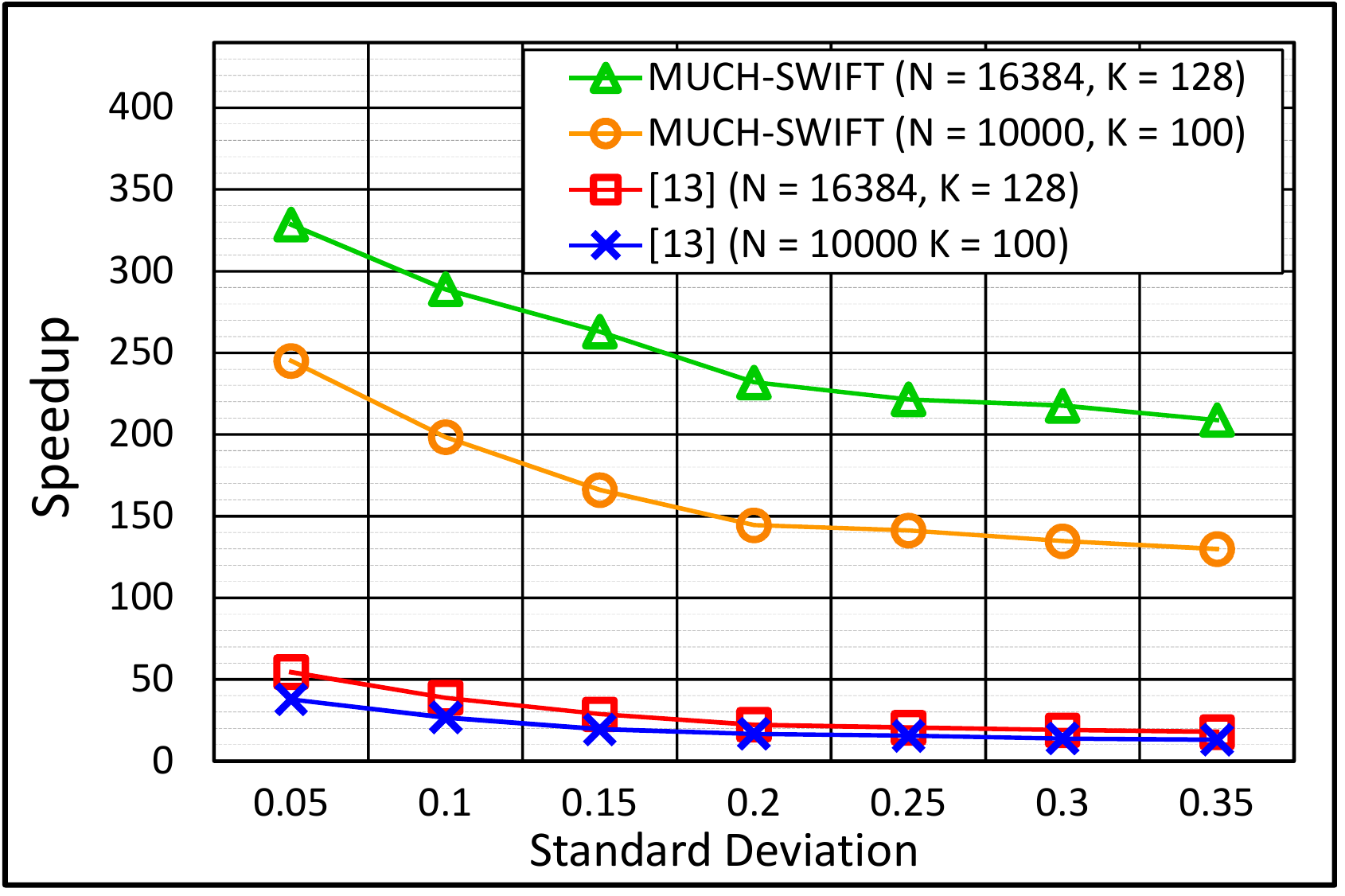} }} \\
    \vspace{-12pt}
    \caption{(a) Average Clock Cycles in each iteration (b) Speedup Against Conventional FPGA-based Single Core }%
    \label{kdtree}%
    \vspace{-19pt}
\end{figure}

As mentioned earlier, an FPGA-based implementation for the filtering algorithm was implemented successfully in \cite{Winterstein2013}. Compared to this architecture, MUCH-SWIFT is a multi-core architecture to implement a parallel structure for the filtering algorithm. Additionally, the two-layer filtering approach provides better throughput. Fig. \ref{kdtree}a illustrates average clock cycles for each iteration in MUCH-SWIFT against \cite{Winterstein2013}. As it can be seen, the multi-core architecture provides around $8.5\times$ speedup on average. Also, as it can be seen in Fig. \ref{kdtree}b, in comparison with an FPGA-based architecture without optimization, it is able to achieve more than $210\times$ acceleration on average against conventional FPGA-based implementation. Although four parallel cores have been employed to divide each dataset into four sub-datasets, it is able to achieve around $8.5\times$ speedup in comparison with a single core filtering algorithm \cite{Winterstein2013}. This result proves the impact of two-layer filtering algorithm. Since, the dataset has been divided into four sub-datasets, not only the computations have been divided into four parallel k-clustering algorithm, the extent of the computations have also mitigated. So, it is able to achieve higher efficiency than the expected results (expected close to $4\times$  speedup). Note that the second level of filtering algorithm will be converged in few iterations, because the outputs of the first level of filtering algorithm is very close to the output after convergence. So, it has little impact on the results. Similar to \cite{Winterstein2013}, the test case is generated with normal distribution with varying standard deviation, and all centroids are distributed between data points uniformly. Also, note that all data communications (interaction) between the host and FPGA, which is accomplished via PCIe interface, are counted for timing evaluation.

Fig. \ref{kdtree}a illustrates the average number of needed clock cycles for each iteration in MUCH-SWIFT against an FPGA-based kd-tree implementation \cite{Winterstein2013}. As illustrated, the MUCH-SWIFT architecture provides around $8.5\times$ speedup on average. Additionally, as reported in Fig. \ref{kdtree}b, MUCH-SWIFT is able to provide up to $330\times$ speed-up compared to an FPGA-based architecture without optimization. Also, it achieves more than $210\times$ acceleration on average compared to an FPGA-based architecture without optimization. Note that all data communications (interaction) between the host and FPGA, which is accomplished via PCIe interface, are counted for timing evaluation. The MUCH-SWIFT's robust and scalable data transfer and DMA management contribute to the reported speedup. In fact, this is why the MUCH-SWIFT achieves $8.5\times$ speedup, compared with a single core filtering algorithm \cite{Winterstein2013} when only utilizing 4 parallel cores as the computation is no longer memory bound. 

\begin{figure}[t]%
    \centering
    \subfloat[]{{\includegraphics[width=240pt]{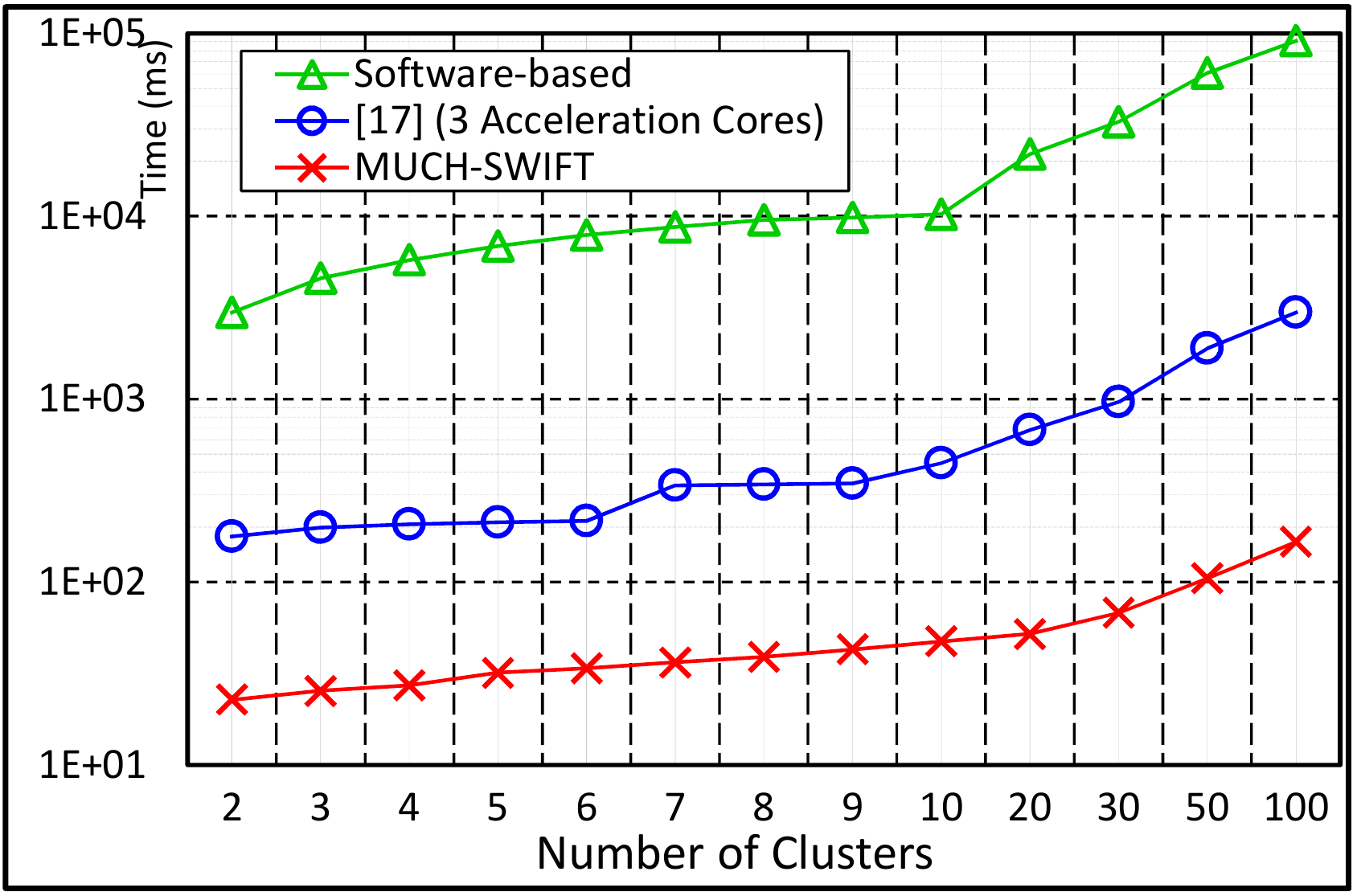} }} \\
    \subfloat[]{{\includegraphics[width=240pt]{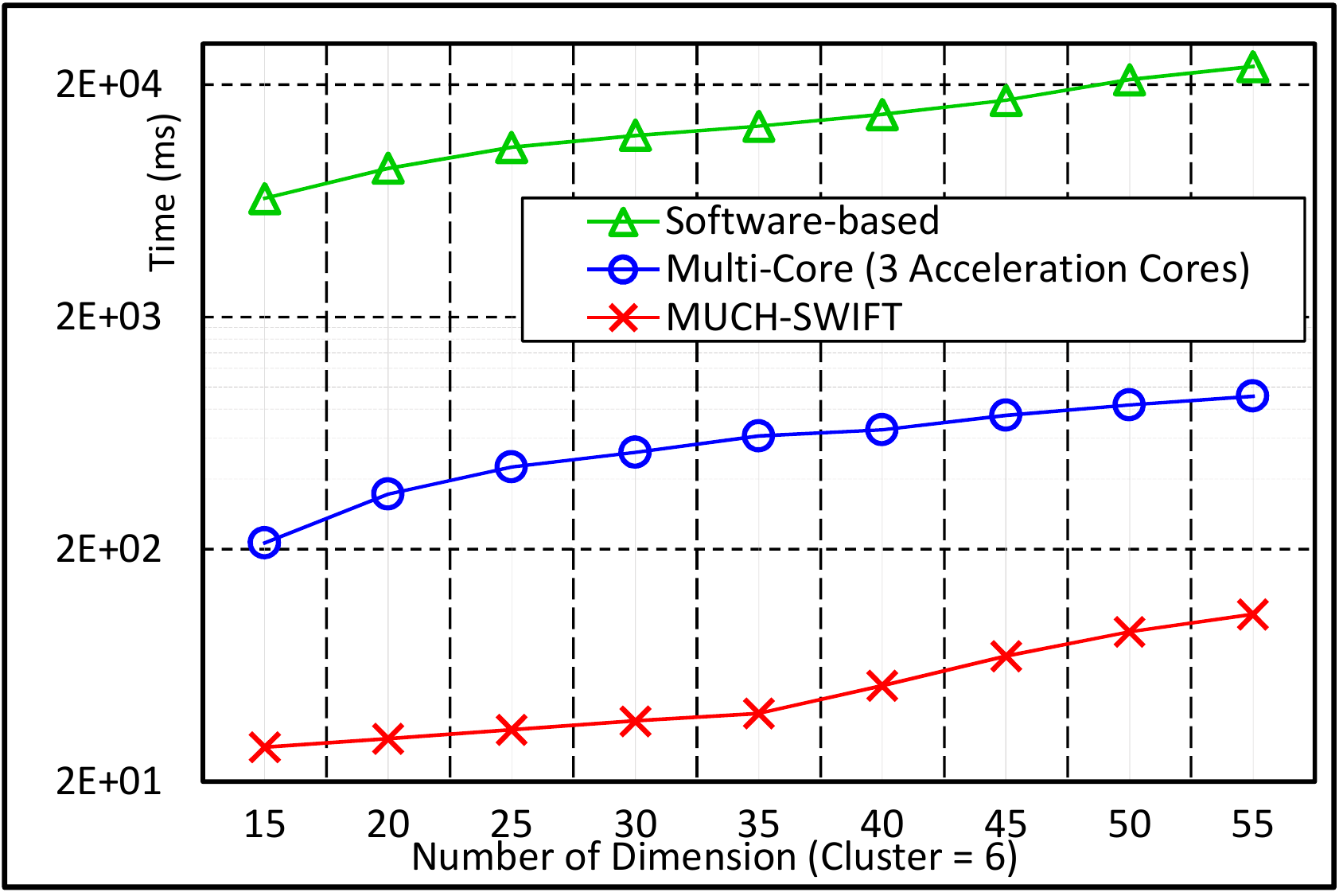} }} \\
    \vspace{-12pt}
    \caption{(a) Execution Time for $10^6$ Data Points (a) with Different Clusters (15 Dimensions) (b) with Different Dimensions (6 Clusters)}%
    \label{multicoreres}%
\end{figure}

In order to illustrate the efficiency of the filtering algorithm with a parallel architecture in comparison with k-clustering implementation without optimization, the MUCH-SWIFT results have been compared with the proposed architecture in \cite{Canilho2016}, which is a multi-core implementation of k-clustering. Fig. \ref{multicoreres}a depicts the execution time of MUCH-SWIFT and \cite{Canilho2016} on 106 data points with 15 dimensions and different number of centroids ranging from 2 up to 100. It is obvious that increasing the number of clusters increases the gap between MUCH-SWIFT and \cite{Canilho2016} due to parallel arithmetic cores in MUCH-SWIFT architecture. In fact, since the number of parallel arithmetic cores in MUCH-SWIFT depends on the number of clusters, and maximum feasible resources on FPGA has been used, it encountered less throughput degradation. Fig. \ref{multicoreres}b focuses on data dimensionality. Fig. \ref{multicoreres} shows around $12\times$ speedup against \cite{Canilho2016} on average.

Table \ref{resource} reports MUCH-SWIFT's resource utilization with the different number of clusters. When increasing the number of clusters, it needs more resources for parallelism, and the available resources on FPGA are limited. So, there is a limit for the number of clusters on FPGA for fully parallel architecture. As reported in Table \ref{resource}, the maximum number of clusters (for fully parallel architecture) is $20$, and for the applications with more clusters, it has to share the parallel modules between clusters uniformly. Note that, $20$ clusters means that it is able to implement $20\times4=80$ parallel modules on ZU9EG, which is significantly large. Also, during implementation phase, the highest proportion of BRAMs and DSPs has been used in order to maximize the number of parallel arithmetic cores.  

\begin{table}[t]
\centering
\caption{Resource Utilization with Different Cluster Sizes}
\vspace{-9pt}
\includegraphics[width = 230pt]{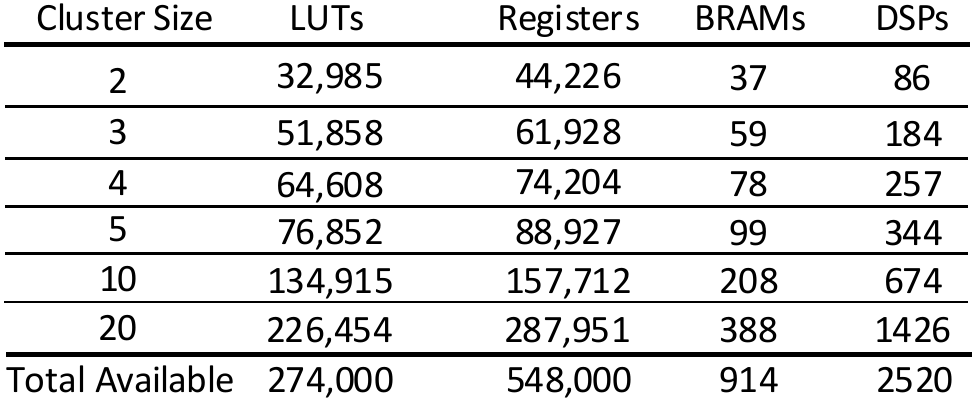}
\label{resource}
\vspace{-15pt}
\end{table} 

\vspace{-5pt}
\section{Related Work}

A HW/SW co-design architecture is implemented in \cite{Gokhale2003} based on NIOS 1.1. But, the HW/SW interface is Peripheral Bus Module (PBM), whose serial infrastructure considerably limits the throughput. Unlike a HW/SW architecture, pure FPGA-based designs \cite{Hussain2011_1, Hussain2011_2} provide significant speed-up against HW/SW co-designs by using fixed-point arithmetic. But, on-chip FPGA memories (like BRAMs) is a big restriction for storing large datasets.

In order to avoid redundant distance calculations, triangle inequality \cite{Elkan2003} has been implemented successfully in \cite{Lin2012}. However, the size of data points is truncated to 8 bits, which is small. A filtering algorithm by using a kd-tree structure is implemented in \cite{Winterstein2013}. Due to using on-chip memories for storing data, it is only able to store 64K data simultaneously. Also, the size of data points is limited to 16, and all computations are based on fixed-point arithmetic.

Another pure FPGA-based k-means clustering architecture is implemented in \cite{Kutty2013}. The number of clusters is fixed in this architecture, and changing it needs re-synthesis and re-implementation. A computer cluster which consists of multiple FPGA-CPU pairs is implemented in \cite{Choi2014}. In this architecture, map-reduce programming is designed to allow easy scaling and parallelization across the distributed computer system, Although this evaluation on multiple FPGA-CPU pairs shows considerable throughput against baseline software implementation, it needs evaluating the case for utilizing multiple FPGAs in processing larger datasets.

A specific FPGA accelerator for the Intel QuickAssist FPGA platform is implemented in \cite{Abdelrahman2016}, which provides an integration between threads in a CPU and an Accelerated Function Unit (AFU) in QuickAssist FPGA. Although, the integration between threads in CPU and FPGA accelerator helps achieving considerable performance, it is applicable only for this specific type of FPGA platform, i.e. IntelAssist. Finally \cite{Canilho2016} presents a ZYNQ-based HW/SW co-design architecture, which employs ARM processors to provide parallelism in both FPGA and ARM processor, but the implemented algorithm has no optimization.

\vspace{-5pt}
\section{Conclusion}

In this paper, we demonstrate that using a HW/SW co-design architecture with a software-based technique provides the maximum efficiency in k-means algorithm. \emph{MUCH-SWIFT}, as an FPGA-based architecture for parallelization of the k-clustering algorithm, has been integrated with a modified two-layer filtering optimization. The MUCH-SWIFT employs all processing cores in the ZYNQ Ultrascale+ SoC to reduce the computation time and a two-layer filtering algorithm designed for parallel processing of binary kd-tree structures. Furthermore, by employing ZYNQ Ultrascale+ and utilizing its DDR3 memory, MUCH-SWIFT increases the feasible size of its input datasets. Additionally, MUCH-SWIFT benefits from the proposed HW/SW co-design architecture, which provides a high-throughput DMA-based PCIe channel for transceiving datasets between the host and ZYNQ SoC. By using this HW/SW co-design architecture, the MUCH-SWIFT achieves around $330\times$ speedup compared to a software-only solution.

\bibliographystyle{ACM-Reference-Format}

\begin{thebibliography}{1}

\bibitem{Jain2011}
Anil~K.~Jain,
"Data Clustering: 50 Years beyond K-means,"
in \emph{Pattern Recognition Letters (PRL)}, Vol. 31, no. 8, pp. 651-666, 2009.

\bibitem{Theiler1997}
J.~P.~Theiler and G.~Gisler,
"Contiguity-Enhanced K-Means Clustering Algorithm for Unsupervised Multispectral Image Segmentation,"
in \emph{Proc. Optical Sci., Eng. and Instrumentation (SPIE)}, pp. 108-118, 1997.

\bibitem{Lloyd1982}
S.~Lloyd,
"Least Squares Quantization in PCM,"
in \emph{IEEE Transactions on Information Theory}, vol. 28, no. 2, pp. 129-137, 1982.

\bibitem{Tan2013}
P.-N.~Tan, M.~Steinbach and V. Kumar,
"Introduction to Data Mining: Pearson New International Edition,"
\emph{Pearson Education Limited}, 2013.

\bibitem{Wu2012}
J.~Wu,
"Advances in K-Means Clustering,"
\emph{Springer-Verlag Berlin Heidelberg}, 2012.

\bibitem{Estlick2001}
M.~Estlick, M.~Leeser, J.~Theiler, and J. J. Szymanski,
"Algorithmic Transformations in the Implementation of K-means Clustering on Reconfigurable Hardware,"
in \emph{Proc. ACM/SIGDA Ninth International Symposium on Field Programmable Gate Array (FPGA)}, pp.103-110, 2001.

\bibitem{Kanungo2002}
T.~Kanungo, D.~M.~Mount, N.~S.~Netanyahu, C.~D.~Piatko, R.~Silverman, and A.~Y.~Wu,
"An Efficient K-Means Clustering Algorithm: Analysis and Implementation,"
in \emph{IEEE Transactions on Pattern Analysis and Machine Intelligence (TPAMI)}, vol. 24, no. 7, pp. 881-892, 2002.

\bibitem{Elkan2003}
C.~Elkan,
"Using the Triangle Inequality to Accelerate K-Means,"
in \emph{Proc. Int'l Conference on Machine Learning (ICML)}, pp. 147-153, 2003.

\bibitem{Gokhale2003}
M.~Gokhale, J.~Frigo, K.~Mccabe, J.~Theiler, C.~Wolinski, and D.~Lavenier,
"Experience with a Hybrid Processor: K-means Clustering,"
in \emph{Journal of Supercomputing}, vol. 26, no. 2, pp. 131-148, 2003.

\bibitem{Sayadi2018aspdac}
H. Sayadi, D. Pathak, I. Savidis and H. Homayoun,
“Power conversion efficiency-aware mapping of multithreaded applications on heterogeneous architectures: A comprehensive parameter tuning,"
in Asia and South Pacific Design Automation Conference (ASP-DAC), pp. 70-75, 2018.

\bibitem{Sayadi2018dac}
H. Sayadi, N. Patel, S. M. P D, A. Sasan, S. Rafatirad, and H. Homayoun,
“Ensemble learning for effective run-time hardware-based malware detection: a comprehensive analysis and classification,"
in Proceedings of the 55th Annual Design Automation Conference (DAC), pp. 1-6, 2018.

\bibitem{Hussain2011_1}
H.~M.~Hussain, K.~Benkrid, H.~Seker, and A.~T.~Erdogan,
"FPGA Implementation of K-means Algorithm for Bioinformatics Application: An Accelerated Approach to Clustering Microarray Data,"
in \emph{NASA/ESA Conf. on Adaptive Hardware and Systems (AHS)}, pp. 248-255, 2011.

\bibitem{Winterstein2013}
F.~Winterstein, S.~Bayliss, and G.~A.~Constantinides,
"FPGA-based K-Means Clustering using Tree-based Data Structures,"
in \emph{Int'l Conf. on Field Programmable Logic and Applications (FPL)}, pp. 1-6, 2013.

\bibitem{Hussain2011_2}
H.~M.~Hussain, K.~Benkrid, A.~T.~Erdogan, and H.~Seker,
"Highly Parameterized K-means Clustering on FPGAs: Comparative Results with GPPs and GPUs,"
in \emph{Int'l Conference on Reconfigurable Computing and FPGAs (ReConFig)}, pp. 475-480, 2011.

\bibitem{Lin2012}
Z.~Lin, C.~Lo, and P.~Chow,
"K-Means Implementation on FPGA for High-Dimensional Data using Triangle Inequality,"
in \emph{Int'l Conference on Field Programmable Logic and Applications (FPL)}, pp. 437-442, 2012.

\bibitem{Sayadi2018cf}
H. Sayadi, S. M. P D, A. Houmansadr, S. Rafatirad, and H. Homayoun,
"Comprehensive assessment of run-time hardware-supported malware detection using general and ensemble learning," 
in ACM Int'l Conf. on Comp. Frontiers (CF), 2018.

\bibitem{Canilho2016}
J.~Canilho, M.~Vestias, and H.~Neto,
"Multi-Core for K-Means Clustering on FPGA,"
in \emph{Int'l Conf. on Field Programmable Logic and Applications (FPL)}, pp. 1-4, 2016.

\bibitem{Sayadi2017iccd}
H. Sayadi, N. Patel, A. Sasan, and H. Homayoun, 
"Machine Learning-Based Approaches for Energy-Efficiency Prediction and Scheduling in Composite Cores Architectures," 
in IEEE Int'l Conf. on Comp. Design (ICCD), pp. 129-136, 2017.

\bibitem{Kutty2013}
J.~S.~S.~Kutty, F.~Boussaid, and A.~Amira,,
"High Speed Configurable FPGA Architecture for K-Mean Clustering,"
in \emph{Proc. Int'l Symposium on Circuits and Systems (ISCAS)}, pp. 1801-1804, 2013.

\bibitem{kamali2018ducnoc}
H. M. Kamali, K. Z. Azar and S. Hessabi, 
"DuCNoC: A High-Throughput FPGA-Based NoC Simulator Using Dual-Clock Lightweight Router Micro-Architecture," 
in IEEE Transactions on Computers, vol. 67, no. 2, pp. 208-221, 2018.

\bibitem{kamali2016adapnoc}
H. M. Kamali and S. Hessabi, 
"AdapNoC: A fast and flexible FPGA-based NoC simulator," 
in 26th International Conference on Field Programmable Logic and Applications (FPL), pp. 1-8, 2016.

\bibitem{Choi2014}
Y.~M.~Choi and H.~K.~H.~So,
"Map-Reduce Processing of K-Means Algorithm with FPGA-Accelerated Computer Cluster,"
in \emph{Proc. IEEE Int'l Conf. on ASAP}, pp. 9-16, 2014.

\bibitem{Abdelrahman2016}
T.~S.~Abdelrahman,
"Accelerating K-Means Clustering on a Tightly-Coupled Processor-FPGA Heterogeneous System,"
in \emph{Proc. IEEE Int'l Conference on Application-Specific Systems, Architectures and Processors (ASAP)}, pp. 176-181, 2016.

\bibitem{Roshanisefat2018srclock}
S. Roshanisefat, H. M. Kamali and A. Sasan, 
"SRCLock: SAT-Resistant Cyclic Logic Locking for Protecting the Hardware," 
in Proceedings of the 2018 on Great Lakes Symposium on VLSI (GLSVLSI), pp. 153-158, 2018.

\bibitem{kamali2016aes}
H. M. Kamali, S. Hessabi,
"A Fault Tolerant Parallelism Approach for Implementing High-Throughput Pipelined Advanced Encryption Standard,"
in \emph{Journal of Circuits, Systems and Computers (JCSC)}, vol. 25, no. 9, 1650113 (1-14), 2016.

\bibitem{kamali2018lutlock}
H. M. Kamali, K. Z. Azar, K. Gaj, H. Homayoun, and A. Sasan, 
"LUT-Lock: A Novel LUT-based Logic Obfuscation for FPGA-Bitstream and ASIC-Hardware Protection," 
in IEEE Computer Society Annual Symposium on VLSI (ISVLSI), pp. 1-6, 2018.

\bibitem{kamali2018swift}
H. M. Kamali and A. Sasan, 
"MUCH-SWIFT: A High-Throughput Multi-Core HW/SW Co-design K-means Clustering Architecture," 
in Proc. of the 2018 on Great Lakes Symp. on VLSI (GLSVLSI), pp. 459-462, 2018.

\bibitem{zynq7000}
Xilinx Inc., 
"Zynq-7000 All Programmable SoC, Technical Reference Manual (UG585),"
[Online]. Available: \newline \seqsplit{https://www.xilinx.com/support/documentation/user\_guides/ug585-Zynq-7000-TRM.pdf}, 2016.

\bibitem{Sayadi2014dft}
H. Sayadi, H. Farbeh, A. M. H. Monazzah, and S. G. Miremadi, 
"A data recomputation approach for reliability improvement of scratchpad memory in embedded systems," 
in IEEE International Symposium on Defect and Fault Tolerance in VLSI and Nanotechnology Systems (DFT), pp. 228-233, 2014.

\bibitem{Sayadi2017igsc}
H. Sayadi and H. Homayoun,
"Scheduling multithreaded applications onto heterogeneous composite cores architecture," 
in 8th International Green and Sustainable Computing Conference (IGSC), pp. 1-8, 2017.

\bibitem{Freund2016}
Karl Freund (Xilinx Inc.), 
"Reconfigurable Acceleration Stack Targets Machine Learning, Data Analytics, and Video Streaming,"
[Online]. Available: \newline \seqsplit{https://www.xilinx.com/support/documentation/white\_papers/acceleration-stack.pdf}, 2016.

\bibitem{Roshanisefat2018bench}
S. Roshanisefat, H.K. Thirumala, K. Gaj, H. Homayoun and A. Sasan,
"Benchmarking the Capabilities and Limitations of SAT Solvers in Defeating Obfuscation Schemes,"
in IEEE 24th Int'l On-Line Testing Symp. (IOLTS), pp. 1-6, 2018.

\bibitem{Wilson2014}
R. Wilson (Altera Inc.), 
"Heterogeneous Computing Meets the Data Center,"
[Online]. Available: \newline \seqsplit{https://www.altera.com/solutions/technology/system-design/articles/2014/heterogeneous-computing.html}, 2014.




\end{thebibliography}

\end{document}